\documentclass[11pt, a4paper]{article}
\usepackage{tikz}
\usetikzlibrary{snakes}
\usepackage[utf8]{inputenc} 
\usepackage{textcomp} 
\usepackage{flafter}  

\usepackage{amsmath,amssymb}  
\usepackage{bm}  
\usepackage{authblk}
\usepackage{memhfixc}  
\usepackage{listings}
\newcommand{\be}{\begin{equation}}
\newcommand{\ee}{\end{equation}}

\def\XXint#1#2#3{{\setbox0=\hbox{$#1{#2#3}{\int}$}
\vcenter{\hbox{$#2#3$}}\kern-.5\wd0}}

\numberwithin{equation}{section}
\begin{document}
\title{Note on an integral expression for the \\ average lifetime of the bound state in 2D}
\author{Thorsten Pr\"ustel} 
\author{Martin Meier-Schellersheim} 
\affil{Laboratory of Systems Biology\\National Institute of Allergy and Infectious Diseases\\National Institutes of Health}
\maketitle
\let\oldthefootnote\thefootnote 
\renewcommand{\thefootnote}{\fnsymbol{footnote}} 
\footnotetext[1]{Email: prustelt@niaid.nih.gov, mms@niaid.nih.gov} 
\let\thefootnote\oldthefootnote 
\abstract
{
Recently, an exact Green's function of the diffusion equation for a pair of spherical interacting particles in two dimensions subject to a backreaction boundary condition 
was used to derive an exact expression for the average lifetime of the bound state. Here, we show that the corresponding divergent integral may be considered as the formal limit of a Stieltjes transform. Upon analytically calculating the Stieltjes transform one can obtain an exact expression for the finite part of the divergent integral and hence for the average lifetime.
}
\section{Introduction}
In Ref.~\cite{TPMMS_SP:2011} an exact analytical expression for the average lifetime
of the bound state and hence the off-rate in 2D was derived, based on an exact Green's function of the reversible diffusion-influenced reaction 
for an isolated pair in two dimensions \cite{TPMMS_2012JCP}. 
Furthermore, it was shown that the associated integral has to be regularized and that its numerical evaluation suggests the relation
\begin{equation}\label{koff}
\frac{1}{k_{\text{off}}} = \frac{1}{\kappa_{d}} + \frac{\ln 2 - \gamma}{2\pi D}\frac{\kappa_{a}}{\kappa_{d}},
\end{equation}  
where $\kappa_{a}$, $\kappa_{d}$, $D$ and $\gamma$ denote the intrinsic association and dissociation constants, the diffusion constant and Euler's number $\gamma=0.5772156649\ldots$ \cite{abramowitz1964handbook}, respectively. 
More precisely, up to a constant, the (regularized) off-rate  is given by  
\begin{equation}\label{integralOffRate}
k_{\text{off}}^{-1} \propto \int^{\infty}_{1}\frac{f(x)}{x}dx+\int^{1}_{0}\frac{f(x) - f(0)}{x}dx,
\end{equation}
where $f(x)$ is defined by
\begin{equation}\label{defF}
f(x):=\frac{P^{2}(x,1)}{x^{2}}
\end{equation}
and $P(x,1)$ is the function
\begin{eqnarray}\label{P}
P(x, 1) &:=& \frac{\frac{2}{\pi}\tilde{h}}{[\tilde{\alpha}(x)^{2}+\tilde{\beta}(x)^{2}]^{1/2}},\\
\tilde{\alpha}(x) &:=& ( x^{2} - \tilde{\kappa}_{D})J_{1}(x) + \tilde{h}xJ_{0}(x), \\
\tilde{\beta}(x) &:=& ( x^{2} - \tilde{\kappa}_{D})Y_{1}(x) + \tilde{h}xY_{0}(x). 
\end{eqnarray}
$J_{0}, J_{1}, Y_{0}, Y_{1}$ denote the Bessel functions of first and second kind and of zeroth and first order, respectively \cite{abramowitz1964handbook}. Furthermore, the dimensionless constants $\tilde{h}, \tilde{\kappa_{d}}$ are related to the intrinsic association and dissociation constants $\kappa_{a}$ and $\kappa_{d}$ by
\begin{eqnarray}
&&\tilde{h}:=ha:=\frac{\kappa_{a}}{2\pi D}, \\
&&\tilde{\kappa_{D}}:=\kappa_{D}a^{2}:=\frac{\kappa_{d}a^{2}}{D}.
\end{eqnarray}
Here, $a$ refers to the encounter radius.
\section{Stieltjes transform}
Instead of approaching the finite integrals in Eq.~\eqref{integralOffRate} directly, we will consider the full divergent integral as the limiting case of a Stieltjes transform. The Stieltjes transform itself can be expressed in terms of modified Bessel functions. Then, their limiting behavior for small arguments opens the possibility to separate the finite and divergent contributions. In this way, we will derive that the finite part gives indeed Eq.~\eqref{koff}.

Starting point is the observation made in Ref.~\cite{Ismail:1977} that a twofold Laplace transform yields a Stieltjes transform
\begin{equation}
\int^{\infty}_{0}e^{-xu}\int^{\infty}_{0}e^{-u\xi}g(\xi)dud\xi=\int^{\infty}_{0}\frac{g(\xi)}{x+\xi}d\xi,
\end{equation}
where $g(t)$ is an arbitrary sufficiently "well-behaved" function.
This observation was used to show that
\begin{equation}
\frac{K_{\nu}(\sqrt{x})}{\sqrt{x}K_{\nu+1}(\sqrt{x})} = \frac{2}{\pi^{2}}\int^{\infty}_{0}\frac{\xi^{-1}}{x+\xi}\lbrace J^{2}_{\nu+1}(\sqrt{\xi}) + Y^{2}_{\nu+1}(\sqrt{\xi})\rbrace^{-1}d\xi,
\end{equation}
for $\nu \geq -1$ and $x>0$, based on the relation
\begin{equation}
\mathcal{L}\bigg\lbrace\frac{2}{\pi^{2}}\xi^{-1}[ J^{2}_{\nu+1}(\sqrt{\xi}) + Y^{2}_{\nu+1}(\sqrt{\xi})]^{-1}\bigg\rbrace = \mathcal{L}^{-1}\bigg\lbrace \frac{K_{\nu}(\sqrt{\xi})}{\sqrt{\xi}K_{\nu+1}(\sqrt{\xi})} \bigg\rbrace,
\end{equation}
cp. Ref.~\cite{Ismail:1977} and references given therein. Here, $\mathcal{L}, \mathcal{L}^{-1}$ denote the Laplace and inverse Laplace transform, respectively and $K_{\nu}$ refers to the modified Bessel function of second kind and $\nu$th order \cite{abramowitz1964handbook}.

Inspired by these results it is tempting to consider the full divergent integral \cite{TPMMS_SP:2011}
\begin{equation}\label{divInt}
\int^{\infty}_{0}\frac{f(x)}{x} dx
\end{equation}
which gives the off-rate as the limit of a Stieltjes transform. Indeed, one finds \cite{TP:2012}
\begin{equation}\label{master}
\frac{\tilde{h}}{\tilde{\kappa_{D}}}\frac{1}{x}-\frac{\tilde{h}K_{1}(\sqrt{x})}{x[(x+\tilde{\kappa_{D}})K_{1}(\sqrt{x})+\tilde{h}\sqrt{x}K_{0}(\sqrt{x})]} = \frac{2}{\pi^{2}}\tilde{h}^{2}\int^{\infty}_{0}\frac{1}{\xi(\xi + x)}\frac{d\xi}{\tilde{\alpha}(\sqrt{\xi})^{2}+\tilde{\beta}(\sqrt{\xi})^{2}}.
\end{equation}
Now, upon changing the dummy variable $\xi \rightarrow \varphi^{2}$, one notes that the divergent integral Eq.~\eqref{divInt} is formally the limiting case of the obtained Stieltjes transform
\begin{equation}
\int^{\infty}_{0}\frac{f(\varphi)}{\varphi} d\varphi = \lim_{x \rightarrow 0} \frac{4}{\pi^{2}}\tilde{h}^{2}\int^{\infty}_{0}\frac{1}{\varphi(\varphi^{2} + x)}\frac{d\varphi}{\tilde{\alpha}(\varphi)^{2}+\tilde{\beta}(\varphi)^{2}}.
\end{equation}
We emphasize again that both the expression on the lhs and on the rhs diverge. However, we can invoke Eq.~\eqref{master} to study the limit $x\rightarrow 0$ and to extract the exact expression for the finite contribution of the divergent integral. To this end, we employ the expansion of the modified Bessel function suitable for small arguments \cite{abramowitz1964handbook} and arrive for small $x$ at (note that $\ln(C):= \ln(\frac{1}{2}) + \gamma$)
\begin{eqnarray}
\text{lhs of Eq.~\eqref{master}} &=& \frac{\tilde{h}}{\tilde{\kappa_{D}}}\frac{1}{x}-\frac{\tilde{h}}{\tilde{\kappa_{D}}x}\frac{\frac{1}{2}\ln(C\sqrt{x})\sqrt{x} + \frac{1}{\sqrt{x}}+\ldots}{\frac{\sqrt{x}}{\tilde{\kappa_{D}}} + \frac{1}{2}\ln(C\sqrt{x})\sqrt{x} + \frac{1}{\sqrt{x}} - \frac{\tilde{h}}{\tilde{\kappa_{D}}}\ln(C\sqrt{x})\sqrt{x}+\ldots}\nonumber\\
&=& \frac{\tilde{h}}{\tilde{\kappa_{D}}}\frac{1}{x} -\frac{\tilde{h}}{\tilde{\kappa_{D}}x}\bigg[1 -\frac{x}{\tilde{\kappa_{D}}} + \frac{\tilde{h}}{\tilde{\kappa_{D}}}x[\ln(C) + \ln(\sqrt{x})]+\ldots \bigg]\nonumber\\
&=& \frac{\tilde{h}^{2}}{\tilde{\kappa_{D}}^{2}}(\ln(2)-\gamma) + \frac{\tilde{h}}{\tilde{\kappa_{D}}^{2}}  -\frac{\tilde{h}^{2}}{\tilde{\kappa_{D}}^{2}}\ln(\sqrt{x}) + \ldots.
\end{eqnarray}
We see that apart from the expected logarithmic divergence we get a finite contribution which exactly yields upon multiplication with the appropriate factor Eq.~\eqref{koff}, cp. \cite[Eq.~(2.11)]{TPMMS_SP:2011}. 
We will elaborate on the whole issue of the 2D off-rate in a forthcoming publication \cite{TP:2012}.  
\subsection*{Acknowledgments}
This research was supported by the Intramural Research Program of the NIH, National Institute of Allergy and Infectious Diseases. 

We would like to thank Bastian R. Angermann and Frederick Klauschen for stimulating discussions.
\bibliographystyle{plain} 
\bibliography{Off_Rate2D} 
\end{document}